\documentstyle[twocolumn,prl,aps,epsfig]{revtex}
\begin{document}
\draft
\title{
Competing frustration and dilution effects on the antiferrromagnetism in
La$_{2-x}$Sr$_x$Cu$_{1-z}$Zn$_z$O$_4$}
\author{I.~Ya.~Korenblit, Amnon Aharony, and O.~Entin-Wohlman}
\address{School of Physics and Astronomy, Raymond and Beverly Sackler Faculty
of Exact Sciences,\\
Tel Aviv University, Tel Aviv 69978, Israel}
\date{\today}
\maketitle
\begin{abstract}
The combined effects of hole doping and magnetic dilution on 
a lamellar Heisenberg antiferromagnet are studied in
the framework of the frustration model.
Magnetic vacancies are argued to remove some of the frustrating bonds
generated by the holes, thus explaining the increase in the
temperature and concentration ranges exhibiting three dimensional
long range order.
The dependence of the N\'eel temperature on both hole and vacancy
concentrations is derived quantitatively from earlier 
renormalization group calculations for the non--dilute case,
and the
results reproduce experimental data with no new adjustable parameters.

\end{abstract}

Since the discovery of high-$T_c$ superconductors much effort
was invested in the investigation of the effect of dopants  
on the magnetic properties of the parent compounds La$_2$CuO$_4$ 
(LCO) and YBa$_2$Cu$_3$O$_6$ (YBCO). 
It is now well established that even a very small dopant concentration,
which introduces a concentration $x$ of holes into
the CuO$_2$ planes, strongly reduces the N\'eel temperature, $T_N$.
In LCO, doped with strontium or with excess oxygen,
 the antiferromagnetic long range order (AFLRO) disappears at a 
hole concentration $x_c \approx 2\%$
\cite{KBB}, while in YBCO $x_c \approx 3.5\%$
\cite{CAM,NBB}.
In contrast, the effect of Cu dilution by nonmagnetic Zn is much weaker.
Like in percolation, the
AFLRO persists at Zn concentration, $z$, as large as 25 \% \cite{HKP}.

Recently, H\"ucker and coworkers~\cite{HKP}  studied the phase diagram of 
La$_{2-x}$Sr$_x$Cu$_{1-z}$Zn$_z$O$_4$, and found surprising results:
 It appears that the vacancies introduced by Zn doping
weaken the   destructive effect of holes (introduced by the Sr)
on the AFLRO. 
E. g., in a sample with $z= 15 \%$, the 
critical concentration $x_c$ of the holes is approximately 3\%,
i.e. {\it larger} than in vacancy free LCO. 
Also, at $x=0.017$ the N\'eel temperature has a {\it maximum}
as function of $z$, implying a {\it reentrant} transition!
To explain these phenomena, H\"ucker {\it et al.} measured 
the variable range hopping conductivity in their samples (at temperatures
lower than 150 K all samples were insulators), and showed that
Zn doping lowers the localization
radius of the holes. Their qualitative
conclusion was that as the holes become more
``mobile", their influence on $T_N$ increases.
However, 
so far there has been no {\it quantitative}
understanding of the combined dependence
of $T_N$ on both $x$ and $z$.

In this paper we present a quantitative calculation, which reproduces
all the surprising
features of the function $T_N(x,z)$.
Our theory extends an earlier calculation \cite{CKA}, 
which treated the effects of quenched
hole doping on the AFLRO in Sr doped LCO, i. e. calculated $T_N(x,0)$.
The same parameters were then used to reproduce the observed
$T_N(x)$ for the bi-layer
Ca doped YBCO \cite{KAE}.
Here we reproduce the full function $T_N(x,z)$, with practically
no additional adjustable parameters.

Our theory is based on the frustration model \cite{aharony}, which argues that
when a hole is localized on a Cu--O--Cu bond \cite{emery,ER},
it effectively turns the
interaction between the Cu spins strongly
ferromagnetic, causing a canting of the
surrounding Cu moments with an angle which decays with the distance $r$
as $1/r$. The frustrating bond thus acts like a magnetic dipole \cite{aharony,vil}.
As argued in Ref. \onlinecite{CKA}, similar dipolar effects also 
arise when the hole is localized over more than one bond.
The frustration  model also predicted a magnetic spin glass phase
for $x>x_c$ \cite{aharony}, as recently confirmed in detail in
doped LCO and YBCO \cite{NBB,WBE,cho,CBK}.
Furthermore, the model successfully reproduced the local field
distributions observed in NQR experiments \cite{GA}.
In earlier work, Glazman and Ioselevich \cite{glazman}
analyzed the planar non--linear $\sigma$ model with random dipolar
impurities, assuming that the dipole moments are
annealed and expanding in $x/T$. 
In Ref. \onlinecite{CKA} we generalized that analysis, 
treating the dipoles as quenched. The
two calculations coincide to lowest order in $x$, but our 
renormalization group analysis
allows a calculation of $T_N$ all the way down to zero at $x_c$, supplying
a good interpolation between these two limits.

In what follows we summarize that theory, with emphasis on the changes 
necessary for
including the Zn vacancies.
We argue that 
the main effects of the vacancies enter in two related ways.
First, the concentration $z$ of the Zn vacancies renormalizes the concentration of
frustrated bonds; when a Cu ion is missing from (at least) one end of a
``frustrated" bond, then this bond is no longer acting like a ``dipole". The
probability to find a bond without vacancies on both ends
is $(1-z)^2$, and therefore the effective concentration of ``dipolar"
bonds is equal to
\begin{equation}
y=x(1-z)^2.
\label{cond}
\end{equation}
Second, 
when one Cu ion at an end of a hole--doped bond is replaced by
Zn, then the strong antiferromagnetic coupling  between the spins of
the second Cu and of the hole on the oxygen
will form a singlet, which is equivalent
to a magnetic vacancy also on the second Cu. 
Hence the holes increase the number of vacancies,
turning their effective concentration into
\begin{equation}
v=z[1+2x(1-z)].
\label{conv}
\end{equation}
In what follows, we shall concentrate on the regime $x<0.03$, where
the $x$-dependence of $v$ on $x$ is very weak.

Following Ref. \onlinecite{CKA}, we descibe the system
by the Hamiltonian 
\begin{equation}
{\cal H}={\cal H}_{v}+{\cal H}_{d},
\label{HH}
\end{equation}
where ${\cal H}_{v}$ is the non--linear sigma
model (NL$\sigma $M) Hamiltonian in the renormalized classical region \cite{CHN}, 
representing the long wave length
fluctuations of the unit vector
${\bf n}({\bf r})$ of antiferromagnetism.
In the presence of short-range  inhomogeneity, this Hamiltonian 
can be written as
\begin{equation}
{\cal H}_{v }={1\over2}\int d{\bf r}\rho_s({\bf r})\sum_{i,\mu}(\partial_{i}
n_{\mu})^{2}. 
\label{hv}
\end{equation}
Here $i=1,...,d$ and $\mu =1,...,{\cal N}$ run over the spatial
Cartesian components and over the spin components, respectively,
$\partial_{i}\equiv\partial /\partial x_{i}$, and  the 
effective local stiffness $\rho_s({\bf r})$ is a random function.
The spatial fluctuations $\delta\rho_s({\bf r})$ of this function
are $\delta$-correlated:
 $[\delta\rho_s({\bf r})\delta\rho_s({\bf r'})]= K\delta({\bf r}-
{\bf r'})$, where [...] means quenched averages. Simple
A
 power counting arguments show \cite{CKA} that $K$ is irrelevant in the
renormalization group sense.
Therefore, we can replace $\rho_s({\bf r})$ in Eq. (\ref{hv}) by its
quenched average $\rho_s(v) \equiv [\rho_s({\bf r})]$.
A

${\cal H}_{d}$ is constructed \cite{glazman} to reproduce the
dipolar canting of the spins at long distances.
 Denoting by ${\bf a}({\bf r}_{\ell})$ the unit
vector directed along the frustrating bond at ${\bf r}_{\ell}$, and by
$M_l{\bf m}({\bf r}_{\ell})$ the corresponding dipole moment
(where ${\bf m}({\bf r}_{\ell})$ is a unit vector giving the
direction of the dipole, and $M_l$ is its magnitude), we have
\begin{equation}
{\cal H}_d=\rho_s(v)\int d{\bf r}\sum_{i}{\bf f}_{i}
({\bf r})\cdot\partial_{i}{\bf n},\label{hint}
\end{equation}
with
\begin{equation}
{\bf f}_{i}({\bf r})=\sum_{\ell}\delta ({\bf r}-{\bf r}_{\ell})M_{\ell}
a_{i}({\bf r}_{\ell}){\bf m}({\bf r}_{\ell}),\label{f}
\end{equation}
where the sum runs only over doped bonds which frustrate the surrounding
(namely have both Cu ions present).

As argued in Ref. \onlinecite{CKA}, the renormalization group procedure 
generates an effective dipole--dipole interaction between the dipole moments,
$\{{\bf m}({\bf r})\}$, which is mediated via the canted Cu spins.
At low temperature $T$ these moments develop very long ranged spin--glassy
correlations,
and may thus be considered frozen.
Hence, we treat all the variables ${\bf r}_{\ell}$, ${\bf a}({\bf r}_{\ell})$,
and ${\bf m}({\bf r}_{\ell})$ as quenched, and we have
\begin{equation}
\bigl [f_{i\mu}({\bf r})f_{j\nu}({\bf r}')\bigr ]
=\lambda\delta_{\mu\nu}\delta_{ij}
\delta ({\bf r}-{\bf r}').
\label{ff}
\end{equation}
Here $\lambda=Ay$, $A=M^2Q/d$, where $Q=[m_{\mu}^2({\bf r})]$, and 
the effective 
dipole concentration $y$ replaces the parameter $x$ used in Ref. 
\onlinecite{CKA}.

With these assumptions, we have now mapped our problem to that treated in Ref. \onlinecite{CKA}. We can thus take over the results from there,
and $T_N(x,z)$ should be equal to the N\'eel temperature derived
there for hole concentration $y$ and stiffness constant $\rho_s(v)$.

The renormalization group analysis of the Hamiltonian (\ref{HH})
\cite{CKA}
found the two--dimensional
antiferromagnetic correlation length $\xi_{2D}$,
as function of the two parameters $t=T/\rho_s$ and $\lambda$.
The results contain exact exponential factors, which give the leading
behavior, and approximate prefactors.
For doped LCO, the results were given for two separate regimes:
\begin{equation}
\xi_{2D}/a=C_1 \lambda^{0.8} \exp\Bigl(\frac{2\pi}{3\lambda}\Bigr)
\label{xilam}
\end{equation}
for $t<\lambda$, and
\begin{equation}
\xi_{2D}/a=C_2\exp\Bigl(\frac{2\pi}{3\lambda}\bigl[1-
\bigl(1-\frac{\lambda}{t}\bigr)^3\Bigr]\Bigr)
\label{xit}
\end{equation}
for $t>\lambda$. Here, $a=3.8\AA$ is the lattice constant, and the coefficients
$C_1$ and $C_2$ may have a weak dependence on $t$ and on $\lambda$. In
Ref. \onlinecite{CKA} the data on La$_{2-x}$Sr$_x$CuO$_4$
were fully described by the constant values
$C_1=0.74$ and $C_2=0.5$.
 
The three dimensional (3D) N\'eel temperature was then derived from the relation
\begin{equation}
\alpha \xi_{2D}^2 \sim 1,
\label{3d}
\end{equation}
representing the appearance of 3D AFLRO due to the weak relative spin 
anisotropy or the weak relative interplanar exchange coupling, both
contained in the parameter $\alpha$.
Combining Eqs. (\ref{xilam}) and (\ref{3d}) thus yields an
$\alpha$-dependent value for the critical value $\lambda_c$, above which
AFLRO is lost. This value should give the critical line for all
$t<\lambda$.
Using the undoped value $\alpha \sim 10^{-4}$ \cite{KBB},   
Ref. \onlinecite{CKA} estimated $\lambda_c \approx 0.37$. Assuming that
$\alpha$ is independent of either $y$ and $v$, this yields $y_c=\lambda_c/A
\approx 0.019$, where we have used the value $A=20$ found for slightly doped LCO \cite{CKA}.
Combining this with Eq. (\ref{cond}), we thus find
\begin{equation}
x_c \approx \frac{0.019}{(1-z)^2},
\label{xc}
\end{equation}
showing an increase of the antiferromagnetic regime with increasing $z$.
At $z=0.15$, this would predict $x_c \approx 0.026$, slightly lower than the
observed value. This discrepancy could result from various sources.
For example, 
dilution may affect the nearest neighbor exchange energy in the plane, $J$,
more strongly than the interplanar interaction or the anisotropy. This
would imply that $\alpha$ increases with
$z$.

Combining Eqs. (\ref{xit}) and (\ref{3d}), one finds
that for $t>\lambda$, the critical line is given by
\begin{equation}
\frac{t_N(\lambda)}{t_N(0)} \approx \frac{B\lambda}{1-(1-3B\lambda)^{1/3}},
\label{tnc}
\end{equation}
where
\begin{equation}
B=-\frac{1}{4\pi} \ln(\alpha C_2^2) \equiv \frac{1}{t_N(0)}.
\label{B}
\end{equation}
We next look at the dependence of $T_N$ on $x$ for fixed $z$. Ignoring the
weak dependence of $v$ on $x$ in Eq. (\ref{conv}), $\rho_s$ is assumed to depend
only on $z$ (the relative error in $\rho_s$ from neglecting $x$ in Eq.
(\ref{conv}) is less than 3\%, see below). 
In that case, $\rho_s$ drops out of the ratio on the LHS 
in Eq. (\ref{tnc}), which becomes equal to $T_N(x,z)/T_N(0,z)$.
The RHS of that equation now depends only on $\lambda=Ay=Ax(1-z)^2$,
reflecting a universality of the plot of $T_N(x,z)/T_N(0,z)$ versus
the rescaled variable $y=x(1-z)^2$.
Note that this universal plot, which should describe the N\'eel temperature
for many values of $z$, requires no new parameters; all the parameters are 
known from the limit $z=0$. In fact, it is worth noting that the RHS of Eq.
(\ref{tnc}) depends only on the combination $B\lambda$, so that it should also
apply to other lamellar systems with different values of $B$,
resulting from different values of $\alpha$.
\begin{figure}[h]
\epsfig{file=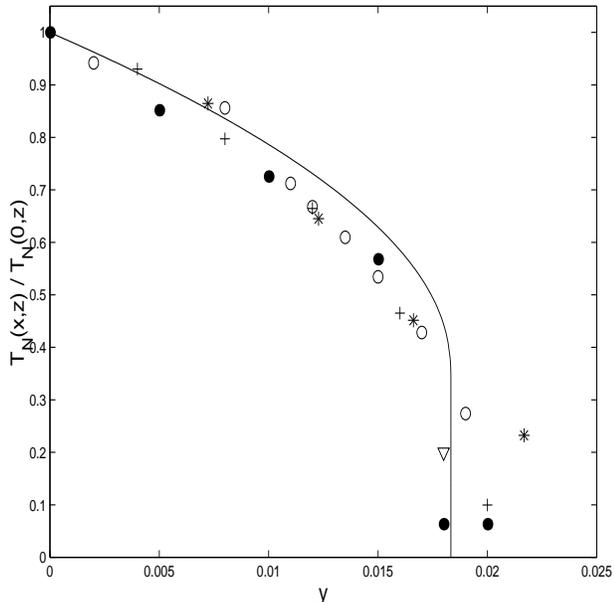,height=8cm,width=8cm}
\caption{The universal plot of $T_N(x,z)/T_N(0,z)$ versus $y$, based on
Eqs. (11) (for $t<\lambda$) and (12) (for
$t>\lambda$).
Symbols
are from experiments: $z=0$: empty circles Ref. [4], full
circles Ref. [19], crosses  Ref. [12],
triangles  Ref. [3]; $z=0.15$: stars Ref. [4].}
\end{figure}
Figure 1 presents the universal plot of $T_N(x,z)/T_N(0,z)$, from both Eqs.
(\ref{tnc}) (for $t>\lambda$) and (\ref{xc}) (for $t<\lambda$).
This theoretical curve is then compared with various experiments, for both
$z=0$ and $z=0.15$. It is satisfactory to note that except for one
point, the data from the latter are indistinguishable from those for
the non--dilute case, confirming our universal prediction.

For comparison of the $z-$dependence of $T_N(x,z)$
with experiments, it is more convenient to scale $T_N(x,z)$
by $T_N(0,0)$. For that purpose, we need the ratio $T_N(0,z)/T_N(0,0)$.
Theoretically, 
Eq. (\ref{B}) yields
\begin{equation}
\frac{T_N(0,z)}{T_N(0,0)}=\frac{t_N(0,z)\rho_s(z)}{t_N(0,0)\rho_s(0)}
=\frac{B(0)}{B(z)}\frac{\rho_s(z)}{\rho_s(0)},
\label{tnzz}
\end{equation}
where the weak $z-$dependence of $B(z)$ may result from such a dependence
of either $\alpha$ or $C_2$ in Eq. (\ref{B}).
According to Refs. \onlinecite{HKP} and \onlinecite{CRS}, the experimental data
 fit the linear dependence
\begin{equation}
\frac{T_N(0,z)}{T_N(0,0)} \approx 1-3.20z.
\label{tnz1}
\end{equation}
up to $z=0.25$.
At low concentrations, $z<0.10$, this is
in good agreement with  the $1/S$ expansion result, \cite{HK}
$\rho_s(z)/\rho_s(0)=1-3.14z$, 
if one uses the approximation $B(z) \approx B(0)$
in Eq. (\ref{tnzz}).
At higher concentrations 
the ratio $\rho_s(z)/\rho_s(0)$ in the classical limit
decreases with dilution approximately as 
$\rho_s(z)/\rho_s(0)=1-3.14z + 1.57z^2$,\cite{HK}
 i.e. slower than the experimental $T_N(z)/T_N(0)$. This discrepancy
can be due to quantum corrections
to $\rho_s$, or to the $z$ dependence of $B$.
Substituting Eqs. (\ref{tnzz}) and
(\ref{tnz1}) (also replacing $z$ by $v$) into Eq. (\ref{tnc}), we have
\begin{equation}
{T_N(x,z)\over T_N(0,0)}=(1-3.20v){ABy\over 1-(1-3ABy)^{1/3}}.
\label{Tn1}
\end{equation}
\begin{figure}
\epsfig{file=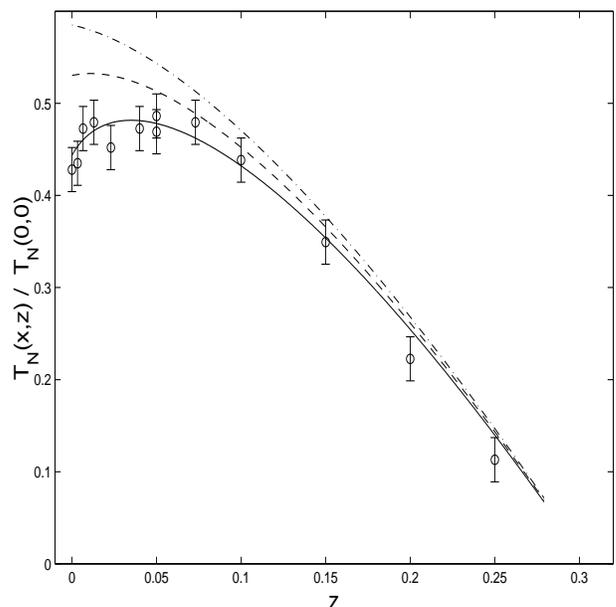,height=8cm,width=8cm}
\caption{$T_N(x,z)/T_N(0,0)$ versus $z$. The full,  dashed, and dash-dotted
line represent Eq. (16), for $x=0.018$, $x=0.017$, and $x=0.016$ respectively.
Circles are experimental data,
from Ref. [4],
for Sr concentration
$0.017\pm 0.001$.}
\end{figure}

Figure 2 shows the dependence of $T_N(x,z)$ on the dilution $z$,
 given by the above equation, 
for three concentrations of holes.
The theoretical curve for $x=0.018$ reproduces very well the 
observed maximum in the dependence of $T_N(x,z)/T_N(0,0)$ on $z$.
The experimental points
were measured at a nominal  Sr concentration $x=0.017\pm0.001$.

An important prediction of the theory is the high sensitivity
of the maximum to the hole concentration. The maximum exists only
at $x$ sufficiently close to $x_c$, and disappears at lower $x$.
 It would be  interesting to check this prediction experimentally.

In conclusion, we found the combined effect of hole doping and magnetic 
dilution on the long-range order in lamellar Heisenberg 
antiferromagnets. We showed that
dilution weakens the destructive effect of the holes on the AFLRO.
The critical concentration $x_c$ increases with dilution,
and the dependence of $T_N$ on vacancy concentration reveals
a maximum, if the hole concentration is sufficiently
close to $x_c$. These findings are in quantitative agreement 
with the experiment.
Furthermore, the experimental data for the two concentrations agree
with each other even better than with the theoretical curve, demonstrating
the validity of the scaling $x \rightarrow x(1-z)^2$ beyond any theory.
We note that the 
consistency of our theory with the experiments, with
no new adjustable parameters, supports the validity of the
frustration model. 

This project has been supported by the US-Israel Binational
Science Foundation.
AA also acknowledges the hospitality of the ITP at UCSB, and
the partial support there for the NSF under grant No. PHY94007194.

\end{document}